\documentclass[iop]{emulateapj-rtx4}

\usepackage{graphicx}  
\usepackage{dcolumn}   
\usepackage{bm}        
\usepackage{amsfonts,amsmath,amssymb,mathrsfs}
\usepackage{color}

\usepackage{hyperref}
\hypersetup{
  colorlinks=true,        
  linkcolor=blue,         
  citecolor=cyan,         
}
\usepackage{natbib,times}
\definecolor{orange}{rgb}{1,0.5,0}

\newcommand{\ie}{\textit{i.e.,}~}
\newcommand{\eg}{\textit{e.g.,}~}

\shorttitle{A novel paradigm for short gamma-ray bursts with extended
  X-ray emission} 
\shortauthors{Rezzolla and Kumar}

\begin{document}

\title{A novel paradigm for short gamma-ray bursts with extended
  X-ray emission}

\author{Luciano Rezzolla\altaffilmark{1}, Pawan Kumar\altaffilmark{2}}

\altaffiltext{1}{Institute for Theoretical Physics,
  Frankfurt am Main, 60438, Germany}
\altaffiltext{2}{University of Texas at Austin, Austin, TX 78712, USA}

\begin{abstract}
The merger of a binary of neutron stars provides natural explanations for
many of the features of short gamma-ray bursts (SGRBs), such as the
generation of a hot torus orbiting a rapidly rotating black hole, which
can then build a magnetic jet and provide the energy reservoir to launch
a relativistic outflow. Yet, this scenario has problems explaining the
recently discovered long-term and sustained X-ray emission associated
with the afterglows of a subclass of SGRBs. We propose a new model that
explains how an X-ray afterglow can be sustained by the product of the
merger and how the X-ray emission is produced before the corresponding
emission in the gamma-band, although it is observed to follow
it. Overall, our paradigm combines in a novel manner a number of
well-established features of the emission in SGRBs and results from
simulations. Because it involves the propagation of an ultra-relativistic
outflow and its interaction with a confining medium, the paradigm also
highlights a unifying phenomenology between short and long GRBs.
\end{abstract}
		
	 	
\section{Introduction}

The merger of a binary system containing at least one neutron star (NS)
represents the most attractive scenario to explain the phenomenology
associated with short gamma-ray bursts (SGRBs), although many
alternatives exist [see~\cite{Berger2013b} for a recent review]. While
merging binaries of neutron stars (BNSs) were suggested already in the
80's~\citep{Narayan92,Eichler89}, numerical simulations
(\citealt{Shibata99d, Baiotti08, Anderson2007, Bernuzzi2011,
  Paschalidis2014}) and new observations~\citep{Berger2013b} have put
this scenario on firmer grounds. In particular, the simulations have
shown that the merger of BNSs inevitably leads to the formation of a
massive metastable object, which can either collapse promptly or survive
up to a fraction of a second emitting large amounts of gravitational
radiation. Furthermore, if the NSs are magnetized, the inspiral can be
accompanied by a precursor electromagnetic
signal~\citep{Palenzuela2013a}, while the merger can lead to
instabilities~\citep{Siegel2013,Kiuchi2014} and to the formation of
magnetically confined jet structures once a torus is formed around the
black hole (BH)~\citep{Rezzolla:2011,Paschalidis2014}.

Despite the progress of simulations, the recent phenomenology of SGRBs
presents a serious riddle for any process involving BNSs. The Swift
satellite~\citep{Gehrels2004} has revealed phases of roughly constant
luminosity in the X-ray afterglows of a large subclass of SGRBs. These
are referred to as ``X-ray plateaus'' (\eg
\citealt{Rowlinson2013,Gompertz2013}) and last $10-10^4\,\mathrm{s}$. The
riddle is then in the timescales involved, which are too long if the
X-ray emission is really an afterglow. The gamma-ray emission, in fact,
is normally associated to an ultra-relativistic jet launched by the BH,
produced by the collapse of the \emph{binary-merger product} (BMP), in
its interaction with the accreting torus. Since, the torus' mass is
$\lesssim0.1\,M_{\odot}$, with accretion rates
$\sim10^{-3}-10^{-2}\,M_{\odot}\,{\rm
  ms}^{-1}$~\citep{Rezzolla:2010,Hotokezaka2013}, the accretion timescale
is at most $\sim1\,{\rm~s}$. This is three or more orders of magnitude
smaller than the observed timescale for the \emph{sustained} X-ray
emission.

A way out from this riddle is in principle available, but it only leads
to a different one. It is possible to invoke the presence a long-lived
``central engine'' in terms of a ``protomagnetar'', that is, a uniformly
rotating object formed in the merger that powers the X-ray emission
through standard dipolar radiation and spin-down~\citep{Zhang2001,
  Metzger2008, Metzger:2011, Bucciantini2012}. Indeed, the BMP can either
be a supramassive NS (a star with mass above the maximum mass for
nonrotating configurations but below the maximum mass for uniformly
rotating configurations) or a BH. The lifetime of the BMP is still very
uncertain, but is likely $\lesssim10^4\,{\rm s}$~\citep{Ravi2014}, which
is long enough to yield a sustained X-ray emission. However, the riddle
in this case is in the timing of the gamma- and X-ray emissions. If the
X-ray emission is produced by the BMP, then it \emph{cannot follow} the
gamma-ray emission, which seems to require a jet and hence a BH. Indeed,
none of the simulations to date indicates the generation of a collimated
jet by the BMP~\citep{Price06, Liu:2008xy, Giacomazzo:2010,
  Palenzuela2013a,Kiuchi2014, Giacomazzo:2014b}, which instead appears
after the formation of a
BH~\citep{Rezzolla:2011,Dionysopoulou2014:inprep}.

\begin{figure*}
  \begin{center}
    \includegraphics[width=7.0cm,height=8.0cm]{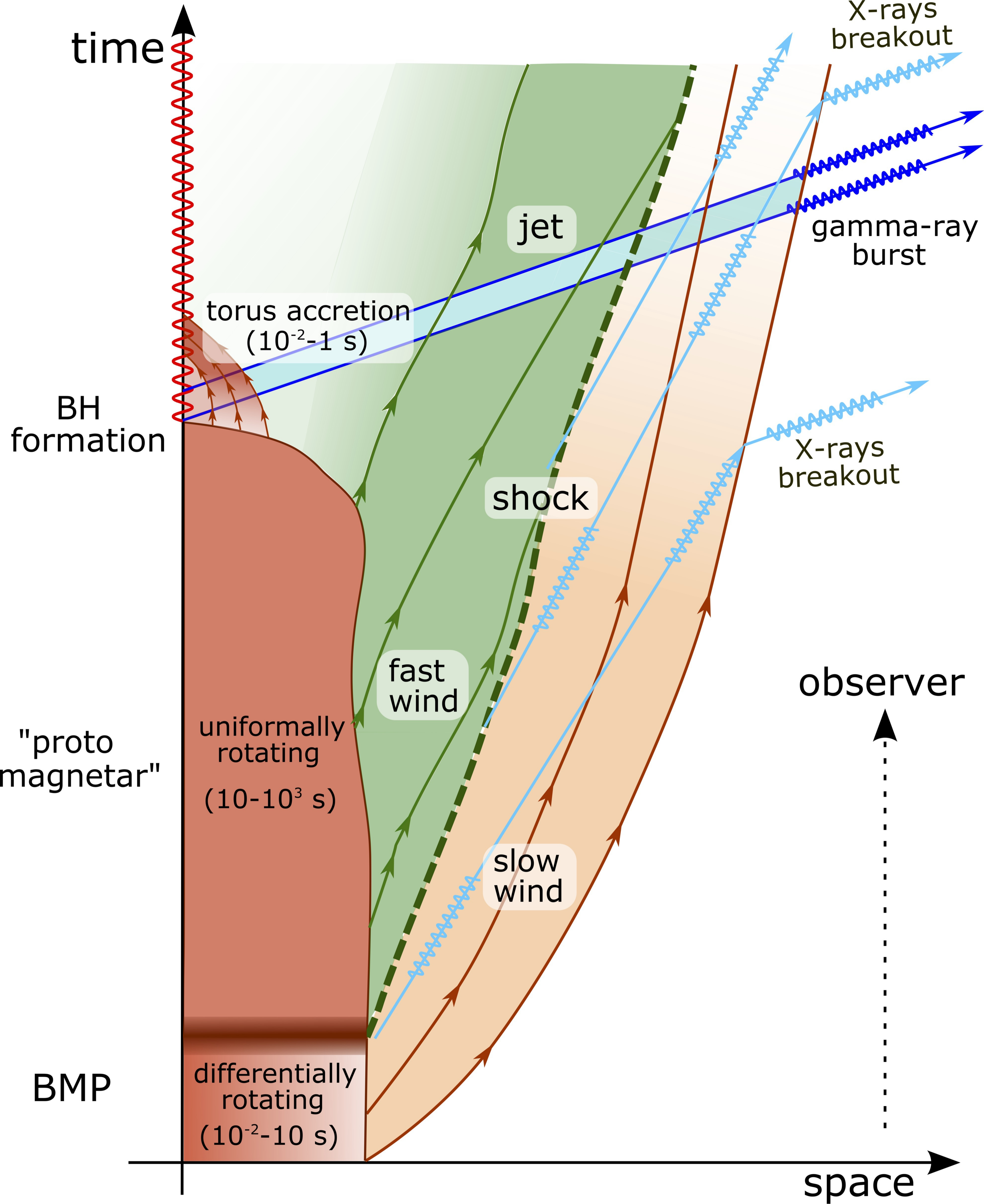}
    \hskip 2.0cm
    \includegraphics[width=7.0cm]{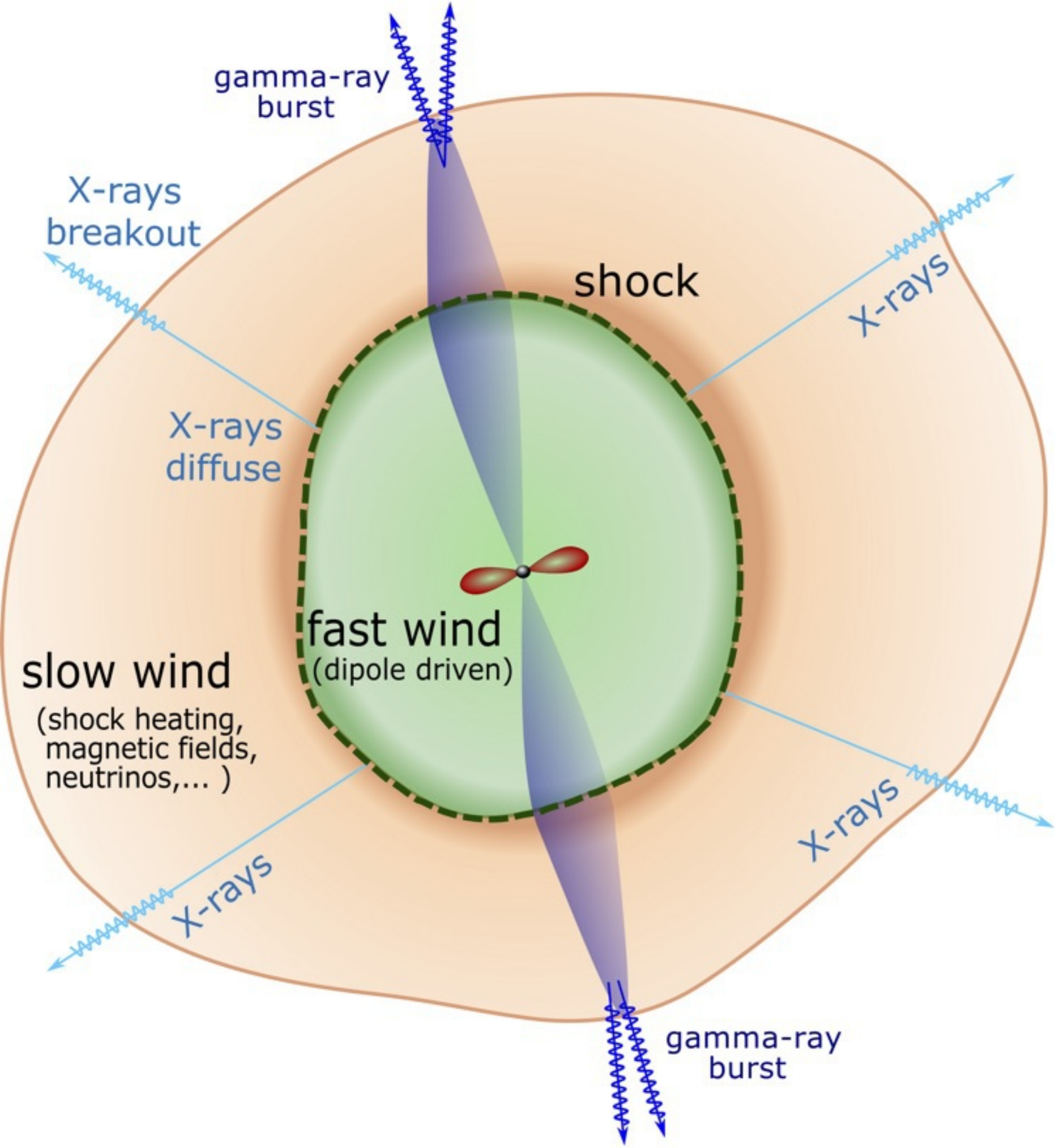}
  \end{center} 
  \caption{\textit{Left:} Schematic spacetime diagram showing in red the
    region occupied by the BMP that eventually collapse leading to
    BH--torus system. Shown in brown and green are the regions occupied
    by the magnetically driven slow wind and by the dipole-driven fast
    wind. The interaction of the two winds generates a shock and the
    sustained X-ray emission, while a jet is produced by the
    BH--torus. \textit{Right:} Schematic snapshot after a BH--torus
    system has been produced and the winds have expanded.}
\label{fig:fig1}
\end{figure*}

\section{The basic picture}

Before diving into the details of our ``two-winds'' model for those GRBs
that have an x-ray afterglow, it is useful to summarize its main
features. The left panel of Fig.~\ref{fig:fig1} presents a spacetime
diagram where shown as red-shaded is the region occupied by the BMP,
which eventually collapses to produce a rapidly rotating BH surrounded by
an accreting torus. The BMP rotates differentially for an Alfv\'en
timescale, \ie $\lesssim1-10\,{\rm s}$ and, assuming it does not collapse
to a BH when differential rotation is lost, it will rotate uniformly for
considerably longer, \ie $\lesssim10^3-10^4\,{\rm s}$. Shown as
brown-shaded is the region occupied by the \emph{slow} and
\emph{baryon-rich} wind, which is approximately spherical and moves at
bulk speeds of $\sim0.01-0.1\,c$ and then progressively slows down as
part of the kinetic energy is lost. This wind can be driven in a number
of ways, possibly all acting at the same time: via shock
heating~\citep{Hotokezaka2013}, via magnetic fields and differential
rotation~\citep{Kiuchi2012b,Franci2013b,Siegel2014}, or via
neutrinos~\citep{Metzger2014,Perego2014}. In all cases, the duration of
the slow wind is $\lesssim1-10\,{\rm s}$, and in the first two scenarios
the wind is isotropic for realistic magnetic-field
topologies~\citep{Siegel2014}, and will be quenched once differential
rotation is suppressed. At this point, the uniformly rotating and
magnetized BMP will emit a \emph{fast} and \emph{baryon-poor} wind
(green-shaded area) moving with bulk speeds of $\sim0.3-0.5\,c$. The BMP
provides a continuous source of dipole radiation over a timescale set by
the stability of the BMP, \ie $\sim1\,{\rm s}-10^3\,{\rm s}$.

Because the slow and fast winds have different velocities, the latter
catches up with the former, producing a shock which heats the matter
locally and leads to an X-ray emission. However, because the matter of
the slow wind is baryon rich and optically thick, the X-ray photons will
not propagate freely, but rather diffuse through the slow-wind material
till reaching a photospheric radius from which they reach the
observer. Because the effective speed of propagation of the X-ray photons
is $\sim\,c/\tau$, where $\tau\gg1$ is the optical depth of the slow wind
where photons are produced, and the shock front moves through the wind
with a relative speed of $\sim\,c/5$, X-ray diffusion can be ignored
until the shock is close to the photosphere.

As the fast and slow winds interact, and the X-ray propagation takes
place through the slow-wind material, the BMP will have spun down via
dipolar emission to a sufficiently slow rate to collapse to a BH
surrounded by a hot dense torus, possibly sending a radio
signal~\citep{Zhang2014,Falcke2013}. Soon after this happens, magnetic
instabilities will develop in the torus, amplifying the magnetic
field~\citep{Rezzolla:2011,Kiuchi2014} and leading to the construction of
a jet-like magnetic structure~\citep{Rezzolla:2011}. This magnetic funnel
can then collimate the low-density material in its interior, which could
be heated either by the neutrinos emitted from the
torus~\citep{Ruffert99b}, or via magnetic reconnection. In addition, the
matter ejected with the slow wind can further confine the propagation of
the jet~\citep{Aloy:2005,Murguia-Berthier2014,Nagakura2014}.  As a
result, an ultrarelativistic jet could be launched propagating with
Lorentz factors $\Gamma\sim100-1000$ (light-blue shaded area). The
dynamics of the jet across the winds material is similar to the one
envisaged for long GRBs, so that a burst of gamma rays is assumed to be
produced as the jet breaks out, with luminosities of
$L\simeq10^{50}-10^{51}\,{\rm erg\,s}^{-1}$, over the timescale of the
duration of the accreting torus, \ie $0.01-1\,{\rm s}$. A snapshot of the
expanded winds is shown in the right panel of Fig.~\ref{fig:fig1}.

In essence, our model solves both the X-ray timescale riddle (the
emission is produced by the BMP, which can survive up to $10^4\,{\rm s}$)
and the timing riddle (the X- and gamma-ray emission are produced at
different times, locations and propagate at different speeds).

\section{Interaction of slow and fast wind and relativistic jet}

A baryon rich slow wind is expected immediately after the merger, which
lasts for a time $t_{sw}\lesssim{1}-10\,{\rm s}$. The mass-loss rate
during this time is $\dot{M}_{sw}\sim10^{-3}\,M_\odot\,{\mathrm s}^{-1}$,
and the wind speed $V_{sw}\sim{c/10}$~\citep{Siegel2014}. The slow-wind
phase ends when the BMP starts to rotate as a solid body, which is also
roughly when the neutrino luminosity drops off, and a global dipole
magnetic field is assumed to emerge~\citep{Zhang2001, Metzger2008}. The
BMP then produces a magnetic-dipole wind with luminosity
\begin{equation}
L_{d}(t)\approx(6\times10^{49}{\rm erg\,s}^{-1})B_{15}^2\,P_{-3}^{-4}\,
\left(1+t/t_{_{\rm SD}}\right)^{-2}\,,
\end{equation}
where
\begin{equation}
t_{_{\rm SD}}\approx(500\,{\rm s})\,B_{15}^{-2}\,P_{-3}^{2}\,,
\end{equation}
is the spin-down time, $B$ is the dipole magnetic field and $P$ is BMP
rotation period. After the neutrino luminosity from the BMP drops, the
dipole wind is expected to have high magnetization, low baryon loading
and relativistic speed. This fast wind drives a shock wave into the slow
wind. The speed of the shock front moving into the slow wind, as seen in
the slow-wind rest-frame, can be obtained from the pressure balance as
viewed in the rest-frame of the shocked wind at radius $r$
\begin{equation}
\frac{L_{d}}{4\pi\,r^2\,c}\approx m_{p}n_{sw}V_s^2\approx\frac{\dot M_{sw}V_s^2}{4\pi\,r^2\,V_{sw}}\,,
\end{equation}
where $V_s$ is the shock speed as seen in the slow-wind rest-frame,
$n_{sw}$ the number density there, and $m_p$ is the unit the baryon
mass. Hence
\begin{equation}
V_s\!\approx\!\left(\frac{L_{d} V_{sw}}{\dot M_{sw}\,c}\right)^{\!\!1/2}\!\!\!\!\!\!\!\approx\!(6\times10^9\,{\rm cm\,s}^{-1})\,L_{d,50}^{1/2}\,\left(\frac{V_{sw}}{c}\right)^{\!\!1/2}\!\!\!\!\!\!\!\sim\!V_{sw}\,.
\end{equation}

The thermodynamic properties of the shocked slow-wind plasma can be
obtained from the Rankine-Hugoniot conditions, which, for a
sub-relativistic shock yield~\citep{Rezzolla_book:2013}
\begin{eqnarray}
\label{eq:rmass}
n_{sw}V_{s}&=&n_{2}V_{2}\,,  \\
\label{eq:momentum}
m_{p}n_{sw}V_{s}^2&=&m_{p}n_{2}V_{2}^2+k_{_{\rm B}}T_{2}n_{2}+\frac{\sigma_{a}T_{2}^4}{3}\,,  \\
\label{eq:energy}
\frac{m_{p} n_{sw} V_{s}^3}{2} &=& \frac{m_{p} n_{2} V_{2}^3}{2}+\left[\frac{5}{2}k_{_{\rm B}}\,T_{2}\,n_{2}+\frac{4}{3}\sigma_a T_{2}^4\right]V_{2}\,,
\end{eqnarray}
where $\sigma_a$ is the radiation constant, $k_{_{\rm B}}$ the Boltzmann
constant, and the index $2$ refers to matter behind the shock. The ratio
of the radiation and gas thermal pressure for the shocked slow wind is
given by
\begin{equation}
\sigma_a T_2^4/(3 k_{_{\rm B}} n_2 T_2) \sim 3 T_2^3/n_{sw}\,,
\end{equation}
When the pressure in the shocked plasma is dominated by the gas thermal
pressure, then we know from the standard shock-jump equations that the
temperature of the shocked gas is $T_2 \sim m_p V_s^2/k_{_{\rm B}}$,
which would be of order 10 MeV (or $10^{11}\,{\rm K}$) for the parameters
of the slow and fast winds. In this case, the ratio of the radiation
pressure to the gas thermal pressure is $\sim 10^{11}$ at $R =
10^{10}\,{\rm cm}$, and the ratio is $10^{13}$ at the shock break-out
radius of $10^{11}\,{\rm cm}$. This large ratio means that the pressure
is completely dominated by radiation, which contradicts the assumption
that the shocked plasma pressure is dominated by gas. So we must drop
this assumption, and consider the opposite situation that the shocked
plasma temperature is dominated by the radiation.  In this case,
Eqs.~(\eqref{eq:rmass})--(\eqref{eq:energy}) yield
\begin{equation}
7\,V_{2}^2-8V_{s}V_{2}+V_{s}^2=0\,,\quad\,{\rm or}\quad\,V_{2}=V_{s}/7\,.
\end{equation}
Substituting this back into (\ref{eq:rmass})--(\ref{eq:momentum}) we find
\begin{equation}
\label{n-shocked-wind}
n_{2}=7\,n_{sw}\,,\qquad{\rm and}\qquad\sigma_aT_{2}^4/3=6m_{p}n_{sw}V_{s}^2/7\,,
\end{equation}
or
\begin{equation}
\label{T-slow-shock}
T_{2}\approx\left(\frac{9\,L_{d}}{14\pi\,c\,\sigma_a\,r^2}\right)^{1/4}\sim{\rm (few)\,keV}\,.
\end{equation}

The X-ray luminosity from the shocked slow-wind when
    the shock emerges above the photosphere is obtained by
    solving photon diffusion equation and is given by
\begin{eqnarray}
L_{X} &\approx& 4\pi\,R_{s}^2\,\sigma_a\,T_{2}^4(\Delta r/t_{sw})\nonumber \\
&\sim&16\pi\,R_{s}\,\sigma_{_{SB}}\,T_{2}^4\,(R_{s}\,\lambda)^{1/2}\,(V_{sw}/c)^{1/2}\,,
\end{eqnarray}
where $\sigma_{_{SB}}$ is the Stefan-Boltzmann constant, while
\hbox{$R_{s}=t_{sw}V_{sw}(V_{s}+V_{sw})/V_{s}\approx t_{sw}V_{sw}$} is
the shock breakout radius, and
\begin{equation}
\Delta r\sim(R_{s}\,\lambda\,c/V_{sw})^{1/2}\,,
\end{equation}
is the radial distance travelled by the photons in $t_{sw}$ with 
mean-free-path
\begin{equation}
\lambda=\frac{4\pi\,R_{s}^2\,m_{p}\,V_{sw}}{\sigma_{_T}\,Y_e\,\dot M_{sw}}\sim10^5\,{\rm cm}\,,
\end{equation}
where $Y_e\approx0.1$ is the electron fraction, $\sigma_{_{T}}$ the
Thomson-scattering cross section, and the numerical value for $\lambda$
is for $\dot M_w=10^{-4}\,M_\odot$s$^{-1}$, $V_{sw} = c/10$ and
$t_{sw}$=30s. Making use of (\ref{T-slow-shock}), we find
\begin{equation}
L_{X}\sim 10^{47}
({\rm erg\,s}^{-1})\,\dot M_{w,-4}^{-1/2}\,R_{s,12}^{1/2}\,L_{d,50}\,.
\end{equation}
This luminosity is in good agreement with that observed in X-ray
plateaus~\citep{Rowlinson2013,Gompertz2013}, and the spectrum is
non-thermal because of Compton scatterings of photons below the
photosphere by electrons accelerated in the shock. A part of
dipole-driven energy could also be emitted in gamma-rays, as seen in some
pulsars with widely ranging efficiencies
($10^{-3}-1$)~\citep{Abdo2013}. After the shock breakout, the temperature
decreases due to adiabatic expansion as $r^{-2/3}$. The luminosity will
decline as $t^{-1/6}$ as long as in the shocked wind $\tau > 1$. At the
shock, Rayleigh-Taylor instabilities may develop~\citep{Blondin:2001} and
particles be accelerated~\citep{Sironi:2011}.

The fast wind from the BMP lasts $t_{fw}\sim10^3-10^4\,{\rm s}$ when the
BMP loses rotational support and collapses to a BH and a torus. An
accurate determination of the mass of this torus is difficult at the
moment, as numerical simulations cannot be performed on these timescales.
Previous general-relativistic simulations of the collapse of cold and
uniformly rotating supramassive stars have indicated that very little
mass is left outside the black hole, with torus masses $\lesssim
10^{-3}\,M_\odot$~\citep{Shibata99e,Baiotti04,Baiotti07}, although larger
masses (\ie $10^{-2}-10^{-1}\,M_\odot$) can be obtained if the equation
of state is sufficiently soft~\citep{Shibata:2003iw}. The reason for this
different behaviour is simple to explain: With a soft equation of state
the BMP will be centrally condensed and its core will collapse more
rapidly than the outer parts, some of which will find themselves on
stable orbits once the BH is formed. Conversely, with a stiff equation of
state, the collapse will be essentially homologous, with the core and the
outer parts collapsing at the same speed and leaving little material
outside the BH's horizon.

Clearly, if the mass in the torus is very small, then it will become
rather difficult to find the energy reservoir needed to launch and
sustain the jet that we expect in our model. However, present simulations
reveal that the angular velocity distribution of the matter in the BMP
has an inner core which is differentially rotating and an outer envelope
that has essentially Keplerian velocities and which effectively behaves
like a ``disk'' surrounding the BMP's core~\citep{Kastaun2014}. The
amount of mass in this disk is large and can be even $30\%$ of the total
rest mass. It is then possible that the dynamics of the inner core of the
BMP and that of the outer layers will be distinct. More specifically, it
is not unreasonable that the inner core loses differential rotation as a
result of magnetic braking on an Alf\'en timescale of $\lesssim1-10\,{\rm
  s}$ and collapses to a BH only much later. On the other hand, because
on Keplerian orbits, the material in the outer layers could be subject to
a magnetorotational instability
\citep{Velikhov1959,Chandrasekhar1960,Balbus1991} and hence behave as a
standard accretion disk onto a rapidly rotating magnetized star, in which
differential rotation does not brake the rotation but transports angular
momentum outwards. Once developed, magnetic turbulence will regulates
accretion, which can either be on the inner core of the BMP or on the BH
once it is is formed. A good fraction of the material in the outer layers
of the BMP would thus remain on quasi-Keplerian orbits on the much longer
viscous accretion timescale, hence leading to the production of a massive
torus around the BH.

Following this line of arguments, which is admittedly qualitative at this
stage, we assume that a relativistic jet is launched a time $t_{fw}$
after the formation of BMP, and it propagates through the fast- and the
shocked-slow wind. Since the fast wind and the jet are moving in the same
direction at high Lorentz factors, very little work is done by the jet to
open a cavity through the fast wind. However, the propagation of the jet
through the slow wind, which has been at least partially shock heated by
the fast wind, can require considerable expenditure of energy. This
interaction also produces a hot cocoon that encapsulates and collimates
the jet. Eventually, both the jet and the hot cocoon rise above the BMP
slow wind surface at a radius
\begin{equation}
\label{Rs}
R_{cj}\sim{t_{fw}}V_{s,sw}\left(1+V_{s,sw}/v_h\right)\,,
\end{equation}
where $V_{s,sw}$ is speed of the slow-wind matter at the time the cocoon
punches through its surface, and $v_h$ is the speed of the jet head as it
moves through the slow wind. Note that $V_{s,sw}\sim{V_{sw}}+V_{s}/7$ if
$R_{cj}\sim{R_s}$, but $V_{s,sw}\sim{V_{fw}}$ if the cocoon breaks out at
a larger radius.

The cocoon undergoes adiabatic expansion after breaking through the BMP
wind and attains mildly relativistic speed, radiating away a part of its
thermal energy in the X-ray band. A far away observer will thus see a
burst of gamma-rays lasting for a time of order the duration of the jet
($\sim{1}$s or less), and a long lasting phase of X-ray emission from the
cocoon \emph{and} from the X-ray photons produced near the shock and
breaking out at the edge of the slow wind. Thermal photons from the
cocoon are also inverse-Compton (IC) scattered by electrons in the jet
and this could give rise to a bright MeV flash.

Following \citet{Ramirez-Ruiz2002}, \citet{Matzner2003}, and 
\citet{Bromberg2011} the pressure
balance between the jet and the wind in the frame comoving with the
jet-head is expressed as
\begin{equation}
\label{energy-flux1}
\rho_{j}c^{2}\Gamma_j^2\Gamma_h^2(V_j-v_h)^2\approx(\rho_{2} c^2+4p_{2})\Gamma_h^{2}v_{h}^{2}+p_{2}\,,
\end{equation}
where $\rho_j$ and $\rho_{2}$ are the densities of the unshocked jet and
the shocked slow-wind, and $\Gamma_j$ and $\Gamma_h$ are the Lorentz
factors of the unshocked jet and the jet-head with respect to the slow
wind. The fast wind compresses the slow wind into a thin shell of radial
width $\sim{R_{cj}/7}$ [Eq.~(\ref{n-shocked-wind})], so that the density
of the shocked slow-wind is
\begin{equation}
\rho_{2}\sim\frac{t_{sw}\,\dot M_{sw}}{4\pi\,R_{cj}^3/7}
\gg \frac{p_2}{c^2}
\,,
\end{equation}
Ignoring $p_{2}$ in (\ref{energy-flux1}), the jet-head speed relative
to the shocked slow-wind medium is
\begin{equation}
\label{beta-h1}
v_h\approx\frac{V_j}{1+\xi_j^{-1/2}}\,,
\end{equation}
where
\begin{equation}
\label{xij}
\xi_j\equiv\frac{4 R_{cj}\,L_j}{7\,\theta_j^2\,c^3\,t_{sw}\,\dot M_{sw}}\,,
\end{equation}
while $L_j$ and $\theta_j$ are jet luminosity and half-opening angle,
respectively. For typical parameters of $t_{sw}\dot
M_{sw}\sim{10}^{-3}\,M_\odot,~\theta_j\sim0.2,~L_j=10^{50},~{\rm
  erg\,s}^{-1}$, and $R_{cj}\sim{10^{10}}\,{\rm cm}$, we deduce that
$\xi_j\approx{3}$ and $v_h\sim{c/2}$.

The cocoon provides collimation for the jet as long as
$\xi_j<\theta_j^{-4/3}$ (\citealt{Meszaros2001, Bromberg2011,
  Bromberg2014}), and, as a result, the jet angle is in general some
function of $r$, which we take to be a power-law.

\begin{figure}
\centering
\includegraphics[width=0.95\columnwidth]{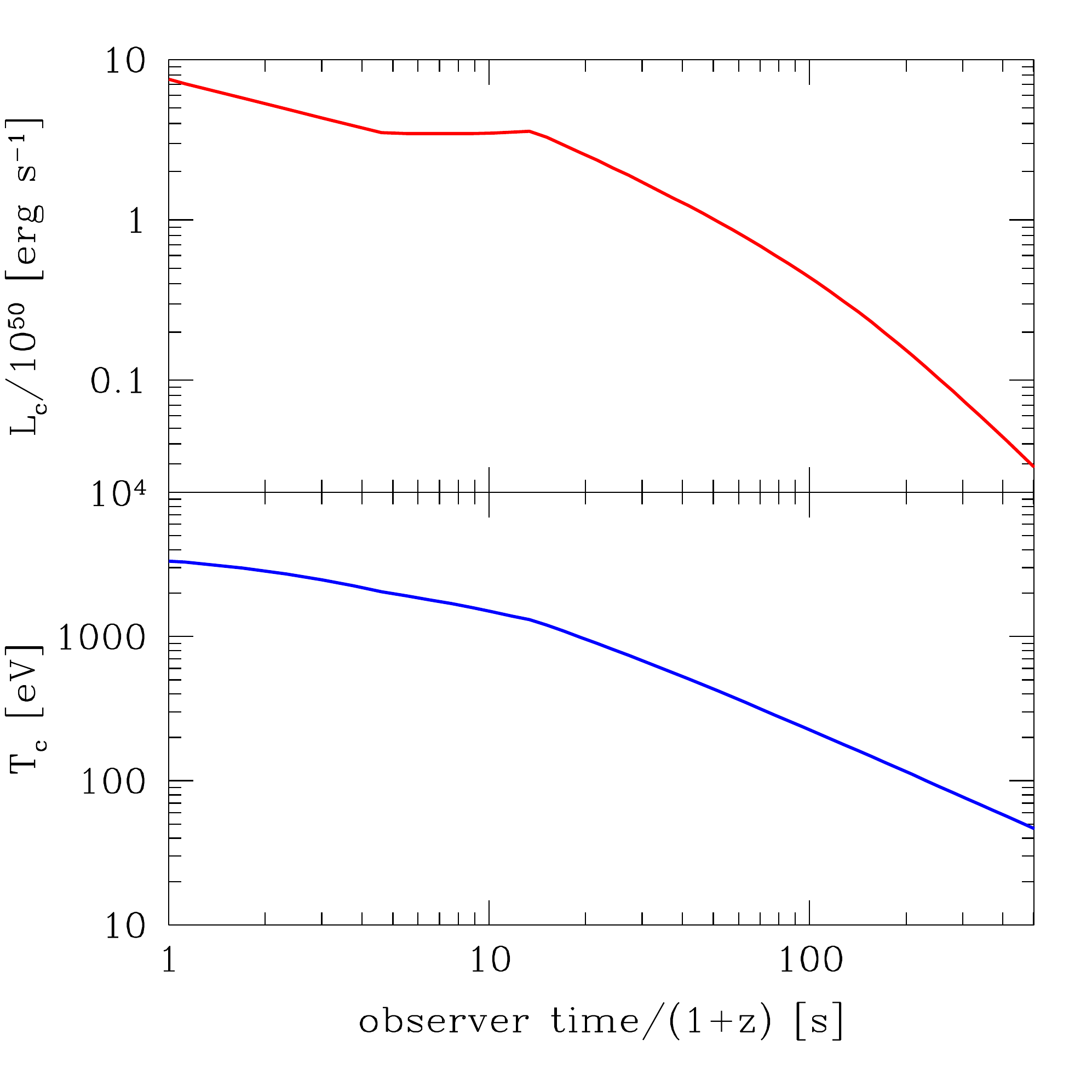}
\caption{The upper panel shows the isotropic equivalent of luminosity of
  cocoon in the host-galaxy frame as a function of observer-frame time,
  where $z$ is the galaxy redshift (We have taken:
  $\dot{M}_{sw}=10^{-3}\,M_\odot\,{\mathrm s}^{-1}$, $V_{sw}=c/2$, and
  $t_{sw}=10$s). The lower panel shows the cocoon temperature.}
 \label{cocoon-prop}
\end{figure}

The energy $E_c$ deposited by the relativistic jet into the cocoon is
roughly equal to the work done by the jet on the wind medium, which is
given by~\citep{Ramirez-Ruiz2002}
\begin{eqnarray}
\!\!\!E_c\approx\!\!\int\frac{dr}{v_h}L_j (1-v_h) 
&\sim&
L_j \left({R_{cj}}/{7c}\right)\nonumber\\
&\sim&
(5\times10^{48}{\rm erg})\,R_{cj,10}\,L_{j,50}\,.
   \label{Ec1}
\end{eqnarray}
The cocoon mass and volume are respectively
\begin{eqnarray}
M_c&\approx&\int{dr\,\pi\,\theta_j^2\,r^2\,\rho_{2}(r)}=\frac{7t_{sw}\,\dot{M}_{sw}}{4}\int{\frac{dr}{r}\theta_j^2}\,,
  \label{m_c}\\
{\cal V}_c &\sim& (R_{cj}/7)^3(v_\perp/v_h)^2\,,
\end{eqnarray}
where
\begin{equation}
v_\perp\sim(p_c/\rho_{2})^{1/2}\,,
\end{equation}
is the transverse expansion speed of the cocoon and $p_c\sim{E_c/{\cal
    V}_c}$ is its average thermal pressure. Combining these equations we
find
\begin{equation}
v_\perp\sim\left[{E_c\,v_h^2\over \rho_{2}}\left({7\over R_{cj}}\right)^3\right]^{1/4}\,.
\end{equation}
Therefore, the cocoon average temperature is given by
\begin{equation}
T_c\approx\left({3 p_c\over \sigma_a}\right)^{1/4}\!\!\!\sim7\times10^8\,{\rm K}\left[{t_{sw}\,\dot M_{w,-4}\,L_{j,50}\over R_{cj,10}^5}\right]^{1/8} v_h^{1/4}\,.
\label{Tc_break}
\end{equation}

Once the cocoon punches through the BMP wind, its Lorentz factor
increases linearly with radius, as is the case for any radiation-dominated
relativistic plasma, until it attains the terminal Lorentz factor  
\begin{equation}
\Gamma_c\approx1+E_c/(M_c c^2)\sim2-10\,,
\end{equation}
at a radius $R_{sat}\approx\Gamma_c\,R_{cj}$. The radiation temperature
in the observer frame during this phase of acceleration does not change
significantly. However, once the cocoon starts to coast at a constant
speed of $v_c$ at $R_{sat}$, the temperature decreases as
$r^{-2/3}$. This decline steepens to $r^{-1}$ for
$r\gtrsim{R}_{cj}\,\Gamma_c^2$, when the radial width of the cocoon
starts to increase linearly with distance.

In the frame comoving with the cocoon, the average number density of
electrons and the photon mean free path are respectively
\begin{equation}
n'_{e,c}(r)\sim\frac{Y_e\,M_c}{m_{p}\,{\cal V}_{c}(r)}\,,\qquad\,\lambda'(r)=\frac{1}{\sigma_T\,n'_{e,c}(r)}\,. 
\end{equation}
The observed luminosity is controlled by photon diffusion below the 
photosphere, and at lab frame time $t$, this is given by
\begin{equation}
\label{L_cocoon}
L_c^{\rm iso}(t)\approx4\pi\,R_{cj}^2\,\sigma_{_{SB}}\,T^4_c(R_{cj})\left(\frac{\lambda'}{ct'+\lambda'}\right)^{1/2}\,,
\end{equation}
as long as $\lambda'(r)\ll r/\Gamma_c$; here $T_c(R_{cj})$ is the cocoon
temperature when it emerges above the BMP-wind surface
[Eq.~(\ref{Tc_break})], while $t'\sim{t/\Gamma_c}$ is the dynamical time
in the cocoon rest-frame. The observed luminosity from the cocoon for
$r\gtrsim\Gamma_c^2\,R_{cj}$ is given by
\begin{equation}
L_c^{\rm iso}(r)\approx4\pi R_{cj}^2\sigma_{_{SB}}T_c^4(R_{cj})\Gamma_c^{4/3}(r/R_{sat})^{-2}(\lambda'/ct')^{1/2}\,.
\label{L_cocoon1}
\end{equation}

The cocoon luminosity and temperature are shown in
Fig.~\ref{cocoon-prop}, with the luminosity being roughly constant until
the radial width of the cocoon starts to increase linearly with time,
which happens at $t\sim1$s for our simplified cocoon model. 

In reality the decline is expected to begin later than shown in
Fig.~\ref{cocoon-prop}, since the cocoon plasma is likely to continue to
escape through the polar cavity at the surface for at least a few times
$R_{cj}/c\sim{t_{fw}}\sim{10^2}\,{\rm s}$, reducing the radial expansion
of the cocoon and flattening the decay of the X-ray lightcurve.  The
decline steepens to $t^{-3}$ when $\tau$ of the cocoon drops below unity,
which could explain the sudden drop-off at the end of plateau in some
SGRBs. Following \citet{Kumar2014}, we find the IC luminosity to be of
order $L_j$, but that lasts only for a short duration of time of order
$R_{cj}/(2 c\Gamma_j^2)$, since the jet is opaque at this radius and
cocoon photons are only scattered when the jet first emerges above the
cocoon photosphere.

\section{Discussion and summary}

The ``two-winds'' model proposed here pieces together events that are
likely to take place when a BNS merges and provides an economical
explanation for several puzzling features in the gamma-ray and X-ray data
for SGRBs. The BMP is expected to be a differentially rotating highly
magnetized object driving a highly baryon-loaded wind with moderate speed
of $\sim0.1\,c$.  This phase lasts for as long as the BMP has substantial
differential rotation, which might be for a few seconds. Subsequently,
the baryon loading decreases, the wind becomes relativistic and is driven
by magnetic-dipole radiation. This fast dipole wind pushes outward the
slow baryon-rich wind and the shock resulting from their interaction
heats up the plasma, generating X-ray radiation with
$L_{X}~\sim{10}^{47}\,{\rm erg\,s}^{-1}$ and lasting for $10^3-10^4\,{\rm
  s}$ with a nearly flat lightcurve. This emission can explain the
puzzling ``plateau'' seen in the X-ray lightcurves of SGRBs. Finally, the
BMP slows down and, no longer being able to resist gravity, collapses to
a BH leaving behind a torus of matter which is accreted onto the BH in
$\lesssim1\,{\rm s}$. How much mass will end up in the torus is hard to
predict at this stage and this represents an obvious weakness of our
model. If the BMP has a stiff equation of state and collapses
homologously, then it is possible that very little matter will be left
outside the BH, making the launching of a jet very problematic if
possible at all. On the other hand, if the BMP has a condensed core
because its equation of state is rather soft, then a non-negligible
amount of matter can be used to build a torus. Clearly, new numerical
simulations are needed to asses this point that may invalidate our model
if it turns out that very little matter is left in the torus.

In the case in which the torus us sufficiently massive, the accreting BH
is expected to produce a relativistic jet responsible for the observed
gamma-ray emission, but the jet first has to make its way out of the
baryon-loaded slow wind. As in long GRBs, the energy required for carving
out a cavity through the wind is converted into thermal energy and
deposited into a cocoon encapsulating the jet and contributing to its
collimation. Because the fast wind has pushed the baryon-rich wind out to
a distance $\sim10^{11}\,{\rm cm}$ by the time a BH forms and jet is
produced, the work done for clearing this cavity is much smaller than
that estimated by \citet{Murguia-Berthier2014}, where only a baryon-rich
wind was considered. A part of the energy deposited by the jet into the
cocoon is radiated away as X-rays when the jet and the cocoon rise above
the wind surface, and could contribute to the extended X-ray
emission. This was already recognized by \citet{Murguia-Berthier2014},
although they did not calculate the emergent radiation as we have in this
work. The emergent X-ray spectra in our models are non-thermal due to the
interaction between photons and electrons accelerated in shocks.

The long lasting X-ray emission described in this work ($\sim10^2\,{\rm
  s}$ duration), arises from two different mechanisms which are: the
shocked slow wind and the hot cocoon surrounding the jet. The X-ray
variability time for the two mechanisms are very different. X-rays from
the cocoon have a short-time variability since the cocoon has a Lorentz
factor of order a few, and its structure is expected to be highly
irregular. The timescale for X-ray luminosity from the cocoon to vary is
of order $\delta t \sim R_{cj}/(2 c \Gamma_c^2)$ which is $\lesssim
1\,{\rm s}$ since $R_{cj} \sim t_{fw} V_{sw} \sim 10^{11}\,{\rm cm}$. The
X-ray luminosity of the cocoon is larger than the luminosity of the
shocked slow wind on a timescale of a few hundred seconds, which is the
duration of observed plateau for short-GRBs, and hence the observed
variability of X-rays of $\sim 1$s is most likely due to the cocoon
emission.  The variably of the X-ray luminosity from the shocked wind
occurs on a longer timescale and is due to the patchiness of the material
of the slow-wind. The density distribution in the slow wind is
approximately spherical but not homogeneous and the shock going through
it produces a hot plasma with density and temperature fluctuations. If
fluctuations occur in the wind on a characteristic lengthscale of
$\ell_f$, then the timescale for fluctuations in the emergent X-ray
lightcurve is $\sim \min\left\{\ell_f^2/(c\lambda),t_d\right\}$, and the
dimensionless amplitude of fluctuations is $\sim (\ell_f/R_s)$; $\lambda$
is photon mean free path and $t_d$ is dynamical time.

While numerical simulations are still too expensive to reproduce
self-consistently this scenario, there are a number of observational
features that can be used to confirm or rule out this novel
paradigm. First, it is clear that in our model the launching of the jet
will take place considerably after the actual merger of the two neutron
stars, which is also when the gravitational-wave amplitude reaches its
first maximum. Hence, the observation of a GRB which is seen to take
place $10^3-10^4\,{\rm s}$ after the maximum gravitational-wave emission
would be a confirmation of the validity of this scenario for SGRBs with
extended X-ray emission.  Second, future observations should be able to
test our model by looking for IC scattered thermal cocoon photons
that should show up at energies $>10\,{\rm MeV}$ with a luminosity
$\sim10^{50}\,{\rm erg\,s}^{-1}$ lasting for about a second. Finally, the
detection of an X-ray emission anticipating the GRB would also represent
a strong validation of this model; indeed the precursor signals in some
SGRBs~\citep{Troja2010} seem already to suggest this possibility.

\medskip
After submitting this paper we learned about the work
of~\citet{Ciolfi2014}, who concentrate on showing that the X-ray photon
diffusion timescale is comparable or larger than the afterglow
timescale. This work on the other hand considers the interaction between
slow and fast winds, creation and propagation of a cocoon by the
relativistic jet passing though the winds, and provides a detailed
calculation of the resulting X-ray lightcurves.  In November 2013, LR and
the authors had a discussion about the timing riddle, but the work has
been developed entirely independently.

\bigskip
\noindent We are grateful to N. Bucciantini, B. Metzger, T. Piran, and
B. Zhang for useful discussions and comments. Support comes from the DFG
Grant SFB/Transregio~7, from ``NewCompStar'', COST Action MP1304, and
from HIC for FAIR.

\bibliographystyle{apj}

\end{document}